\documentstyle[prl,aps,multicol,preprint]{revtex} 
\tightenlines
\newcommand{\beq}{\begin{equation}}
\newcommand{\eeq}{\end{equation}}
\newcommand{\bea}{\begin{eqnarray}}
\newcommand{\eea}{\end{eqnarray}}
\begin{document}
\draft

\title{Anisotropic Condensation of Helium in Nanotube Bundles} 
\author {M. W.  Cole$^1$, V. H. Crespi$^1$, G. Stan$^1$, J. M. Hartman$^1$,   
S. Moroni$^2$, and M. Boninsegni$^3$} 
\address{ 
$^1$Department of Physics and Center for Materials Physics, 
104 Davey Laboratory,\\ Pennsylvania State University, University
Park, PA 16802-6300, USA\\ 
$^2$INFM, Universita' di Roma ``La Sapienza'', Roma, Italy\\ 
$^3$Department of
Physics, San Diego State University, San Diego, CA 92182, USA
\\
} \date{\today}
\maketitle
\begin{abstract}
Helium atoms are strongly attracted to the interstitial channels
within a bundle of carbon nanotubes.  The strong corrugation of the
axial potential within a channel can produce a lattice gas system
where the weak mutual attraction between atoms in neighboring channels
of a bundle induces condensation into a remarkably anisotropic phase
with very low binding energy. We estimate the binding energy and
critical temperature for $^4$He in this novel quasi-one-dimensional
condensed state. At low temperatures, the specific heat of the
adsorbate phase (fewer than 2\% of the total number of atoms) greatly
exceeds that of the host material.
\end{abstract}

\begin{multicols}{2}

Low temperature research on Helium was initially stimulated by the
challenge of determining the condensation temperature of bulk
He\cite{onnes}. In recent decades, two-dimensional He films, in which
the superfluid transition differs qualitatively from that of the
bulk\cite{kosterlitz}, have been particularly intriguing. While once
of only academic interest\cite{liniger,lieb,shulz,takahashi}, Helium
in one-dimensional or quasi-one-dimensional systems has received
increased attention recently since the realization that such systems
can be created in the laboratory.  He atoms are very strongly bound
within the hexagonal lattice of narrow interstitial channels between
tubes within the triangular lattice of a bundle of carbon
nanotubes\cite{Iijima,bundle,stan3,teizer}. Within this very narrow
channel, the transverse degrees of freedom are frozen out even at
relatively high temperatures of $\sim$50 K. The binding energy per
atom, 340 K, is the highest known for He, almost twice that calculated
for He within individual nanotubes\cite{stan1} and 2.4 times higher
than that on the basal plane of graphite\cite{elgin,derry}. It exceeds
the ground state binding energy of bulk liquid $^4$He by nearly fifty
times\cite{debruyn}.

Here we describe how the strong axial confinement of the He
wavefunctions within a single channel can produce a direct
experimental realization of a lattice gas model, wherein the weak
coupling between atoms in neighboring channels induces a
finite-temperature transition into a remarkably anisotropic and
extremely weakly bound condensed state.  First we present a localized
model wherein the Helium resides in periodic array of relatively deep
potential wells; this ``bumpy channel'' approximation is supported by
single-particle Helium band structure calculations. For comparison, we
also describe a delocalized model which assumes translational
invariance within each channel (a ``smooth channel''
approximation). The large difference between the models in the
transition temperature to the condensed phase demonstrates the
importance of the external potential in controlling the within-channel
He--He interaction.

In all calculations we assume that the He--He interaction is
unaffected by the carbon environment, an approximation that omits
screening by both phonons and electrons. On planar graphite the
electrodynamic screening of the van der Waals interaction is most
important and reduces the well depth of the He--He pair potential by
$\sim$ 10\% for a monolayer film\cite{kern,gottlieb}. The smaller
He--C separation in the interstitial channel of a nanotube bundle
should yield a somewhat larger effect; the omission of this screening
implies a moderate overestimate in the binding energies described
below.

The external potential that a Helium atom experiences due to the
Carbon environment of the interstitial channel has a significant
corrugation.  Using a C--He pair potential to model this
interaction\cite{stan3}, a band structure calculation of an isolated
$^4$He atom within an interstitial channel of a (18,0)
\cite{wrapping-index-definition} tube lattice yields a purely one-dimensional
dispersion with a very large enhancement of the lowest-band effective
mass $m^*$ above the bare mass $m$: $m^*/m \approx 18$, with a
bandwidth of 0.18 K (18 times smaller than the free-particle
bandwidth)\cite{jake}. For a regular lattice of tubes the interchannel
tunnelling is negligible.  Fig. 1 shows contours of constant
probability density for the ground state wavefunction at $k=0$ in the
single-particle band structure. The coupling between sites is so weak
that even at moderately high temperatures the primary means of
intersite motion along a channel is single-particle atomic tunnelling.
Due to the heterogeneity of currently accessible nanotube systems
(i.e. mixtures of tubes with different wrapping angles and diameters),
this result should be treated qualitatively, as a demonstration that
the atomic He states are well-localized axially within the
interstitial channel. Such a small bandwidth implies well-confined
single-particle wavefunctions which, in this particular geometry,
occupy a regular lattice with a separation of 4.2 \AA \ between
sites. (In the many-body problem, the correlation due to hard-core
interactions with neighbors could induce a further localization). Note
that the single-particle calculation described above correctly
predicted the binding energy per atom, as recently measured
experimentally\cite{teizer}. The bandwidth could be controlled by
changing the distribution of nanotube diameters\cite{eklund} or
through external pressure, possibly inducing a quantum phase
transition.

For this lattice of localized Helium sites, the natural description is
a lattice gas model wherein the statistical degrees of freedom at low
energy are the occupation indices of the sites, i.e. 0 or
1\cite{porepore}. The intersite hopping energy (i.e. $\sim$0.1 K) is
significantly lower than the potential energy of interaction between
atoms on neighboring sites in a channel ($\sim$0.5
K)\cite{overlap}. Multiple occupancy is excluded by the hard-core
repulsive interaction between atoms (estimated to impose an energetic
cost of at least 100 K). Intra-site excitations also involve high
energy scales, which are irrelevant at low T.  The calculation uses
results previously obtained\cite{fisher,graim} for an anisotropic
simple cubic Ising model with an interaction strength $J_z$ between
neighboring spins along the $z$ axis (i.e. within the same channel)
and a transverse interaction $J_t =c \ J_z$. For the He--nanotube
system, $c$ is very small. When $c < 0.1$, the transition temperature
is well approximated by the asymptotic formula
\cite{fisher,graim}
\beq
{{\rm T}_{\rm c} \over J_z} =
{2 \over {\rm ln}(1/c) - {\rm ln} [{\rm ln} (1/c)] }. \label{eq:a}
\eeq
Here we have $J_z = |V(a)|/4$, where $V(a)$ is the equilibrium
interatomic interaction at the intersite separation and the factor of
$1/4$ arises from the familiar transformation from the Ising model to
the lattice gas. For Helium with an intrachannel site spacing of $a=4.2$
\AA, $J_z$ = 0.5 K. As the present lattice is honeycomb rather than
simple cubic and as second neighbors are not included in
Eqn. \ref{eq:a}, we estimate that the transverse interaction strength
$J_t$ should be renormalized by roughly a factor of $3 \times
3/4$, which includes a factor of $\sim$3 for the second neighbors in
adjacent channels and a factor of $3/4$ for the reduced coordination
of the lattice. For Helium with interchannel spacing $d=9.8$ \AA\ we
obtain $J_t \approx 7$ mK, and $c
\approx 0.015$. Eqn. (\ref{eq:a}) then yields
\beq
{\rm T}_{\rm c} \approx 0.7 J_z.
\label{eq:c}
\eeq
The transition temperature for condensation in this lattice-gas Helium
model is then T$_{\rm c}\sim 0.36$ K. Reducing $c$ by even a factor of
5 would change T$_{\rm c}$ by less than 25\%, illustrating the
insensitivity of T$_{\rm c}$ to the transverse interaction. Variation
in T$_{\rm c}$ through changes in $|V(d)|$ will be minor so long as
this lattice gas model remains valid. The binding energy of this
system is lower than that of any other stable condensed atomic system,
to our knowledge.

Although this system falls in the three dimensional Ising model class,
the quasi-one-dimensional character of the system manifests itself in
the specific heat above the transition.  For small $c$, the specific
heat above the transition\cite{graim} (shown in Fig. 2 at
the critical site occupancy of 1/2) closely follows the one
dimensional result,
\beq {\rm C(T)/(k_B N)}
= \biggl[{J_z\over {\rm k_B T}} \ {\rm sech}\biggl ({J_z\over {\rm k_B T}} \biggr )
\biggr ]^2
\eeq
The one dimensional specific heat has a maximum of $\approx 0.44$ near
T$/J_z=0.83$, slightly above the transition. We take this temperature
as a convenient reference point for comparison to other systems. For
$d=9.8$ \AA\ and $a=4.2$ \AA, the resulting adsorbed particle density
implies a specific heat of $\sim$ 4 mJ/gK (normalized to the mass of
the carbon host) at $T\approx 0.83 J_z$. This contribution to the
specific heat should be observable: it greatly exceeds the background
specific heat of the host material. For example, measurements on a
sample of single-walled carbon nanotubes yield a specific heat of $C
\sim$ 0.2 mJ/gK at 1 K which is decreasing with decreasing
temperature\cite{mizel}.  The specific heat of graphite is at least
three orders of magnitude smaller in this temperature range
\cite{nicklow}. Note also that experimental measurements of the
specific heat of single-walled nanotube bundles significantly exceed
theoretical estimates for the contributions from the nanotube
substrate\cite{mizel}; the low-energy degrees of freedom of adsorbed
gases might account for this discrepancy.

Comparison of these results to calculations in the smooth channel
approximation, wherein the external potential is flat, reveals the
importance of the confinement to defined lattice sites in this
strongly anisotropic condensed He.  Diffusion Monte Carlo calculations
for a one dimensional assembly of $^4$He atoms in a flat
potential\cite{stan3,boninsegni} yielded a very weakly bound state
($\sim$ 2 mK per atom) of remarkably low density ($\sim$ 0.04
\AA$^{-1}$). We introduce the interchannel interactions
through a variational wave function which is a product of identical
states of density $\rho$ in every interstitial channel. The energy
shift $\Delta$ due to the interchannel interaction between channels
separated by distance $d$ is \beq
\label{integral}
\Delta=\rho \ \int_0^\infty dx \
V(r^\prime)
\eeq
where $r^\prime = \sqrt {x^2 + d^2}$. A straightforward numerical
integration of (\ref{integral}), using the Aziz interatomic potential
for Helium\cite{aziz} and including the three nearest and the six
next-nearest neighbor channels yields $\Delta/\rho = -0.228$ K
\AA. Because of the rapid decay of $V$ with distance, inclusion of
more distant channels produces no appreciable change. Fig. 3 shows how
the interchannel interaction increases the binding energy ($2 \to \
$16 mK) and the equilibrium density ($0.035 \to \ $ 0.080 \
\AA$^{-1}$) of the liquid state above that in the single-channel
picture. 

Classical statistical mechanics implies a proportionality between
the ground state cohesive energy and the critical temperature of a
given system. Although this law of corresponding states fails for
quantum systems, a specific class of systems, (i.e., a definite De
Boer quantum parameter\cite{deboer}), typically has a strong
correlation between the critical temperature and the ground state
cohesive energy. For example, $^4$He in three dimensions has a binding
energy per particle of 7.17 K and a critical temperature 5.2 K. In two
dimensions, these values are 0.87 K and 0.85 K,
respectively\cite{gordillo,whitlock2}. The smooth channel model should
then condense at $\sim$ 10 mK. The variational approximation
underestimates T$_{\rm c}$, whereas the neglect of fluctuations in
this nearly one dimensional system overestimates T$_{\rm c}$.

The delocalized smooth channel approximation yields a much lower
transition temperature ($\sim$ 10 mK) than the localized case
($\sim$ 0.3 K). For the smooth channel, the kinetic energy
maintains a large ($\sim$ 15 \AA) distance between He atoms within a
channel, which reduces their interaction energy by about 30 times
relative to the bumpy channel approximation. In the more realistic
bumpy channel model, the kinetic energy arises mainly from the
curvature of the site-localized single-particle wavefunction and is
not relevant to the condensation temperature. The external axial
potential forces a much smaller He--He separation and thereby produces
a much higher transition temperature to the condensed phase.

In experimentally produced bundles of nanotubes the distribution of
nanotube diameters is rather sharp, but present evidence suggests a
heterogeneous distribution of wrapping angles for the graphene layers
comprising individual tubes. The consequent variations in the binding
and axial separation of adsorbtion sites\cite{quasicrystalline} can
strongly affect the critical phenomena near the transition and the
nature of the condensed phase, but should cause comparatively minor
variations in the transition temperature of condensation. As T$_{\rm
c}$ is linearly proportional to $J_z$ and only weakly dependent on
$J_t$ in this regime, the main effect on T$_{\rm c}$ will arise from
variations in the intra-channel intersite separation. Should the
spatial extent of axial localization be substantially longer in some
bundle geometries, then double occupancy could become important as
well.

In summary, we demonstrate that Helium within the interstitial
channels of carbon nanotube bundles can condense into a novel strongly
anisotropic phase wherein the strong axial confinement induced by the
external potential of the surrounding nanotubes greatly enhances the
density and hence the transition temperature to the condensed
phase. At low temperatures the specific heat of the helium absorbates,
which comprise less than 2\% atomic fraction, is predicted to exceed
greatly that of the much stiffer background material. While we focus
on $^4$He, this intriguing interstitial-channel condensation could be
observable for other small atoms, for example, H$_2$ \cite{H2} and
Ne\cite{stan3}. Preferential adsorption to the interstitial sites can
be guaranteed by choosing small adsorbates which energetically prefer
the tighter coordination of the interstitial channel over the interior
of a tube (or by simply using nanotubes with closed ends). For $^3$He,
we expect the ordering behavior to be very similar to that of $^4$He
(except for the possibility of a very low temperature ($\sim \mu$K)
magnetic transition in the Fermi species), since atomic exchange is
negligibly small.  This situation is reminiscent of the epitaxial
ordering transition of $^3$He and $^4$He on the graphite basal
plane. In that case, exchange is much higher than here, yet transition
temperatures of the isotopes differ by less than
1\%\cite{bretz,schick}.

We are very grateful to Kurt Binder, Moses Chan, Veit Elser,
H.-Y. Kim, James Kurtz, David Landau, and Ari Mizel for helpful
discussions and communications. This research has been supported by
the David and Lucile Packard Foundation, the National Science
Foundation through grants DMR-9705270, DMR-9802803, and DMR-9876232,
the Army Research Office grant number DAAD19-99-1-0167, and the
Petroleum Research Fund of the American Chemical Society through
grants PRF 34778-GB6, PRF 31641-AC5, and PRF 33824-G5. We acknowledge
the National Partnership for Advanced Computational Infrastructure
and the Center for Academic Computing at the Pennsylvania State
University for computational support.

\end{multicols}

\begin{figure}
\label{GroundStateFig}
\caption{Isosurfaces of constant probability density 
for the $k=0$ state in the lowest band for He atoms within the
interstitial channel of a lattice of (18,0) nanotubes. The axis of the
chanel is oriented horizontally. The three surfaces correspond to
probability densities of $10^0$, $10^{-2}$, and $10^{-4}$ \AA$^{-3}$.
99\% of the probability is within the middle surface and 99.98\% lies
within the outer surface. The axial latttice constant is 4.2 \AA. The
lowest-band states are well-confined to distinct lattice sites.}
\end{figure}

\begin{figure}
\label{HeatFigure}
\caption{Specific heat as a function of reduced temperature at one-half
site occupancy in the anisotropic lattice-gas He model with $J_z$ =
0.5 K (thin curve). The specific heat diverges at T$_{\rm c}$ = 0.36 K
and asymptotes to the one dimensional specific heat (dotted curve) at
high temperatures. The lower (thick) curve shows a theoretical result for the
specific heat of the nanotube substrate ($\times$ 1000)\protect\cite{mizel}.}
\end{figure}

\begin{figure}
\label{BindingEnergy}
\caption{Energy $\epsilon$ per $^4$He atom in an isolated smooth
one-dimensional channel (solid line), compared to the energy per atom
for a hexagonal lattice of smooth channels at the same interchannel
separation as for a carbon nanotube bundle (dashed line).}
\end{figure}


\begin{references}
\bibitem{onnes} H. Kamerlingh Onnes, Proc. Roy. Acad. Amsterdam {\bf13},
1903 (1911).

\bibitem{kosterlitz}
J. M. Kosterlitz and D. J. Thouless
in {\sl Progress in Low Temperature Physics VIIB}, ed. by D. F. Brewer,
(Interscience, New York, 1978), pp. 371--335.

\bibitem{liniger}
E. H. Lieb and W. Liniger, Phys. Rev. {\bf 130}, 1605 (1963).

\bibitem{lieb}
E. H. Lieb and D. C. Mattis, eds., {\sl
Mathematical Physics in One Dimension} (Academic Press, New York, 1966),
pp. 395--403.

\bibitem{shulz}
T. Giamarchi and H. J. Shulz, Phys. Rev. B {\bf 37}, 325 (1988).

\bibitem{takahashi}
M. Takahashi, {\sl Thermodynamics of One-Dimensional Solvable Models},
Cambridge University Press (1999).

\bibitem{Iijima} S. Iijima, Nature {\bf 354}, 56 (1991).

\bibitem{bundle} A. Thess, R. Lee, P. Nikolaev, H. Dai, P. Petit, J.
Robert, C. Xu, Y. H. Lee, S. G. Kim, A. G. Rinzler, D. T. Colbert, G. E.
Scuseria, D. Tom{\'a}nek, J. E. Fischer, and R. E. Smalley,
Science {\bf 273}, 483 (1996).

\bibitem{stan3}
G. Stan, M. Boninsegni, V. H. Crespi, and M. W. Cole, J. Low Temp. Phys.
{\bf 113}, 447 (1998).

\bibitem{teizer}
W. Teizer, R. B. Hallock, E. Dujardin, and T. W. Ebbesen, Phys. Rev. Lett.
{\bf 82}, 5305 (1999).

\bibitem{stan1}
G. Stan and M. W. Cole, Surf. Sci. {\bf 395}, 280 (1998).

\bibitem{elgin}
R. L. Elgin and D. L. Goodstein, Phys. Rev. A {\bf 9}, 2657 (1974).

\bibitem{derry}
G. Derry, D. Wesner, W. E. Carlos, and D. R. Frankl, Surf. Sci. {\bf 87},
629 (1979).

\bibitem{debruyn}
R. De Bruyn Ouboter and C. N. Yang, Physica B {\bf 44}, 127 (1987).

\bibitem{kern}
R. Kern and M. Krohn, Physica Status Solidi A {\bf 116}, 23 (1989).

\bibitem{gottlieb}
J. M. Gottlieb and L. W. Bruch, Phys. Rev. B {\bf 48}, 3943 (1993).

\bibitem{wrapping-index-definition}
Tubes are indexed by expressing the circumference vector in lattice
coordinates. see e.g. R. Saito, M. Fujita, G. Dresselhaus, and M. S.
Dresselhaus, Appl. Phys. Lett. {\bf 60}, 2204 (1992).

\bibitem{jake}
J. Hartman, V. H. Crespi, M. Boninsegni and M. W. Cole, unpublished.

\bibitem{eklund}S. Bandow, S. Asaka, Y. Saito, A. M. Rao, L. Grigorian, and
P. C. Eklund, Phys. Rev. Lett. {\bf 80}, 3779 (1998).

\bibitem{porepore}
This model is distinct from the interpore correlations of
(R. Radhakrishnan and K. E. Gubbins, Phys. Rev. Lett. {\bf 79}, 2847
(1997)) since in the current case the kinetic energy can be neglected
to lowest approximation.

\bibitem{overlap}
In other configurations with larger intersite separations the ratio of
the hopping energy (which drops exponentially with separation) to the
potential energy of interaction (which drops as a power law) is likely
to be even smaller.

\bibitem{fisher} M. E. Fisher, Phys.Rev. {\bf 162}, 480 (1967).

\bibitem{graim} T. Graim and D. P. Landau, Phys. Rev. B {\bf 24}, 5156 (1981).

\bibitem{mizel}
A. Mizel, L. X. Benedict, M. L. Cohen, S. G. Louie, A. Zettl, N. K. Budraa,
and W. P. Beyermann, Phys. Rev. B, in press.

\bibitem{nicklow}
R. Nicklow, N. Wakabayashi, and H. G. Smith, Phys. Rev. B {\bf 5}, 4951
(1972). 

\bibitem{boninsegni}
M. Boninsegni and S. Moroni, J. Low Temp. Phys., in press.


\bibitem{aziz}
R. Aziz, V. P. S. Nain, J. S. Carley, W. L. Taylor and G. T. McConville,
{J. Chem. Phys.} {\bf 70}, 4330 (1979).

\bibitem{deboer} J. De Boer, Physica {\bf 14}, 139 (1948).

\bibitem{gordillo} M. C. Gordillo and D. M. Ceperley, Phys. Rev. B {\bf
58}, 6447 (1998).

\bibitem{whitlock2}
P. A. Whitlock, G. V. Chester and B. Krishnamachari, Phys. Rev. B {\bf 58},
8704 (1998).

\bibitem{quasicrystalline}
Quasicrystalline order is possible due to incommensurate unit cells in
constituent tubes.


\bibitem{H2}
M. S. Dresselhaus, K. A. Williams, and P. C. Eklund, unpublished (1999);
Q. Wang, S. R. Challa, D. S. Sholl, and J. K. Johnson, Phys. Rev. Lett.
{\bf 82}, 956 (1999); F. Darkrim and D. Levesque, J. Chem. Phys. {\bf 109},
4981 (1998).


\bibitem{bretz} M. Bretz, Phys. Rev. Lett. {\bf 38},501 (1977).

\bibitem{schick} M. Schick, J. S. Walker and M. Wortis, Phys. Rev. B {\bf
16}, 2205 (1977).

\end{references}
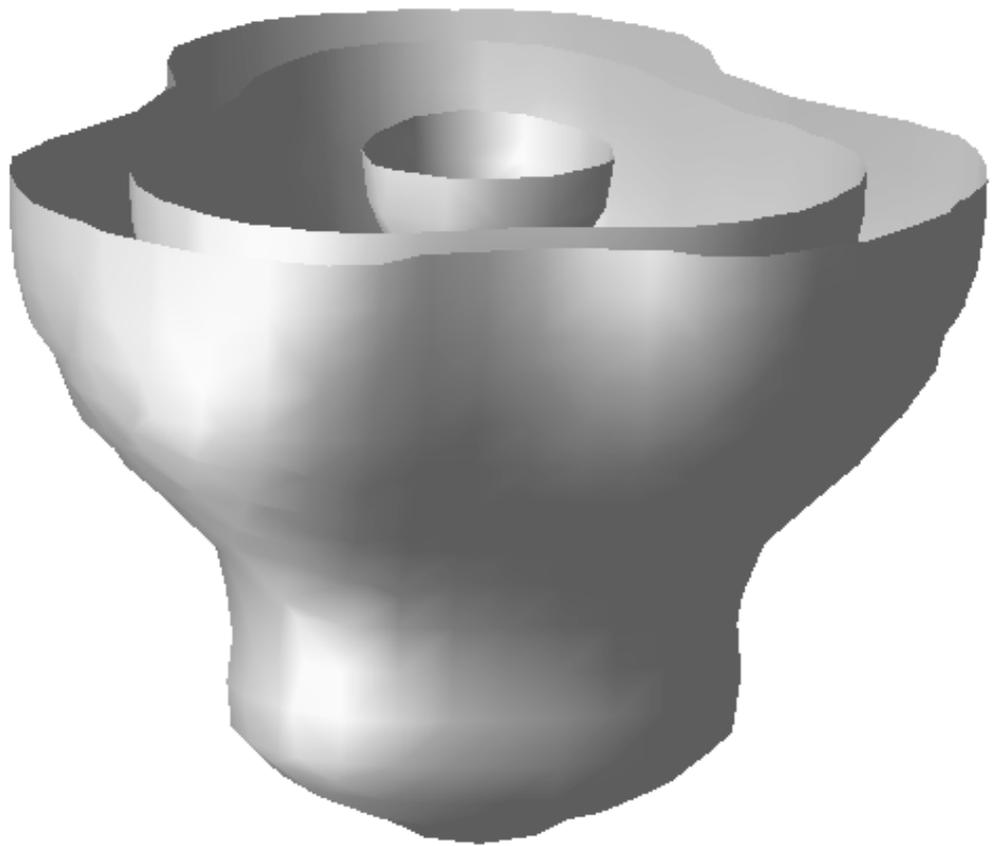
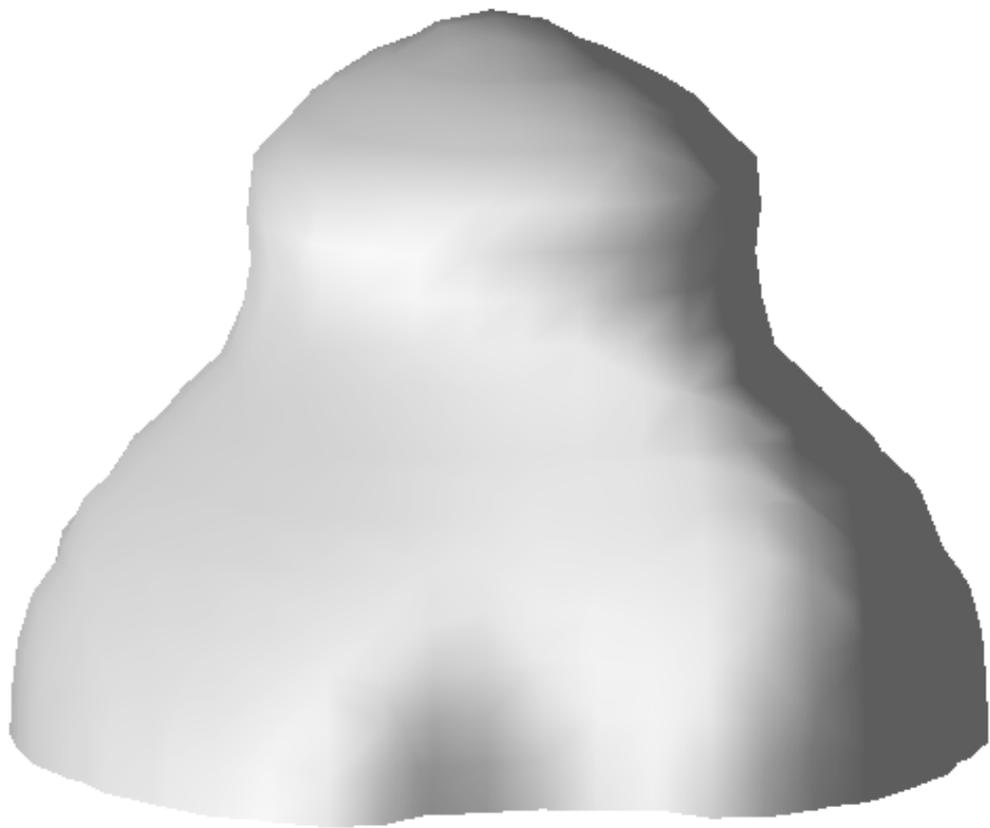
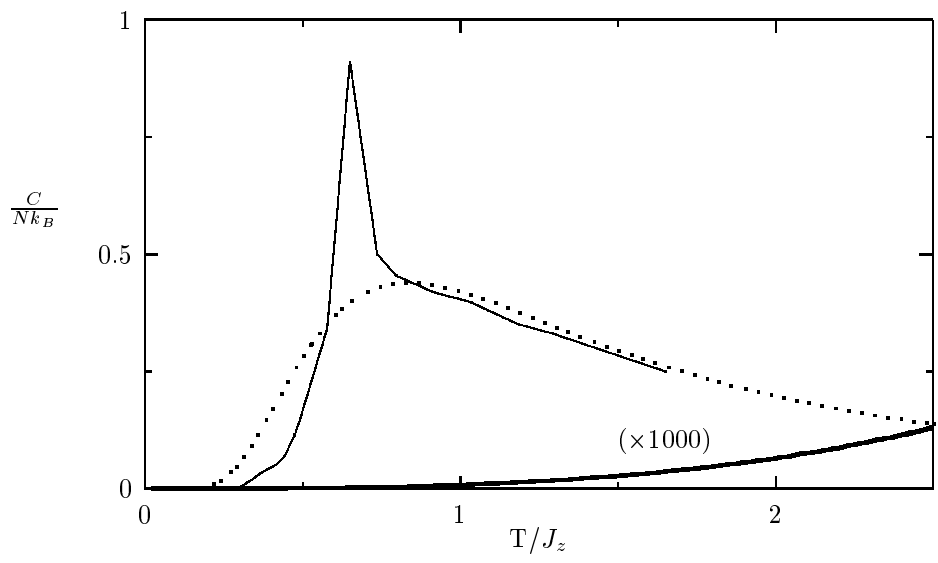
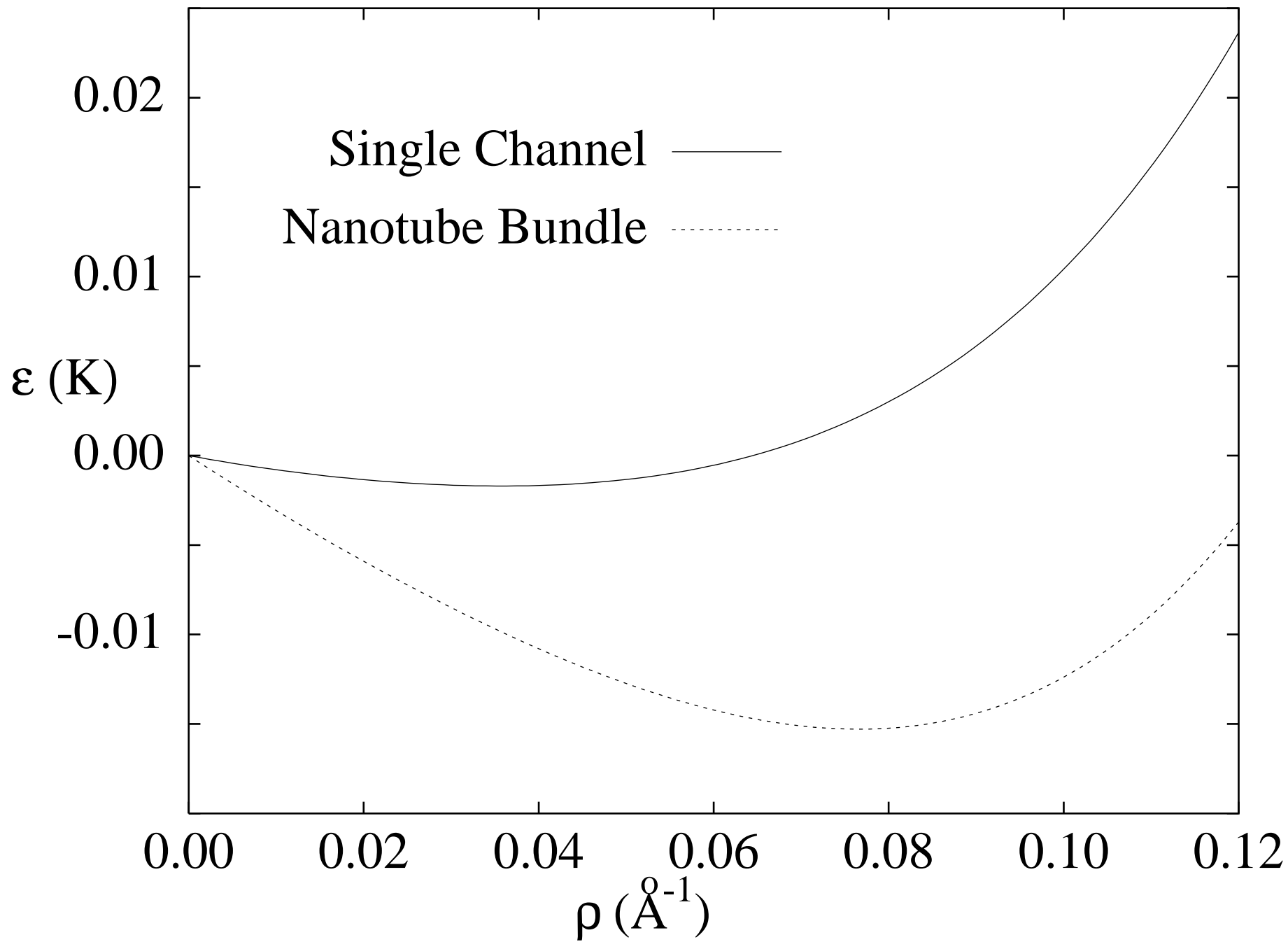
\end{document}